\documentstyle[12pt]{article}
\newlength{\dinwidth}
\newlength{\dinmargin}
\newcommand{\ba}{\begin{array}}
\newcommand{\ea}{\end{array}}
\newcommand{\be}{\begin{equation}}
\newcommand{\ee}{\end{equation}}
\newcommand{\bea}{\begin{eqnarray}}
\newcommand{\eea}{\end{eqnarray}}

\def\bee{\begin{eqnarray}}
\def\eee{\end{eqnarray}}
\def\be{\begin{equation}}
\def\ee{\end{equation}}

\begin{document}
\thispagestyle{empty}
\addtocounter{page}{-1}
\begin{flushright}
IASSNS HEP 98-83\\
SNUTP 98-118\\
{\tt hep-th/9810154}\\
\end{flushright}
\vspace*{1.3cm}
\centerline{\Large \bf Dual D-Brane Actions on 
$AdS_5 \times S^5$~\footnote{
Work supported in part by  the U.S. Department of Energy under Grant No. 
DE-FG02-90-ER40542, KOSEF Interdisciplinary Research Grant and 
SRC-Program, 
Ministry of Education Grant BSRI 97-2418, SNU Faculty Research Grant, 
and 
The Korea Foundation for Advanced Studies Faculty Fellowship. }}
\vspace*{1.2cm} \centerline{\large \bf 
Jaemo Park${}^a$ and Soo-Jong Rey${}^b$}
\vspace*{0.8cm}
\centerline{\large\it School of Natural Sciences, 
Institute for Advanced Study}
\vskip0.3cm
\centerline{\large \it Olden Lane, Princeton NJ 08540 USA${}^a$}
\vspace{0.5cm}
\centerline{\large\it Physics Department \& Center 
for Theoretical Physics}
\vskip0.3cm
\centerline{\large\it Seoul National University, 
Seoul 151-742 KOREA${}^b$}
\vskip0.75cm
\centerline{\tt jaemo@sns.ias.edu, \quad sjrey@gravity.snu.ac.kr}
\vspace*{1.5cm}
\centerline{\Large\bf abstract}
\vspace*{0.5cm}
Utilizing coset superspace approach, dual actions of super 
D1-and D3-branes 
on $AdS_5 \times S^5$ are constructed by carrying out 
duality transformation 
of world-volume $U(1)$ gauge field. Resulting world-volume 
actions are shown 
to possess expected $SL(2,{\bf Z})$ properties. Crucial ingredient for 
deriving $SL(2, {\bf Z})$ transformation property of 
the D-brane actions is 
covariance of $SU(2,2|4)$ coset superspace algebra under $SO(2)$ 
rotation 
between two ten-dimensional Type IIB Majorana-Weyl spinors.

\vspace*{1.1cm}

\baselineskip=20pt
\newpage

\section{Introduction}
Recently, motivated by the Maldacena's conjecture on AdS/CFT 
correspondence, closed forms of $\kappa$-symmetric superstring and 
D3-brane actions have been constructed using the coset superspace 
approach~\cite{MeTs, MeTs2}. 
The formal superspace expression of the superstring or D-brane
actions on Type-IIB supergravity backgrounds have been 
constructed previously
by various authors~\cite{Gri,Ce,BeTo}. However, using the
coset superspace approach, it has become possible to obtain an 
explicit form 
of the action in terms of the coordinate fields $(x, \theta)$.
This development is quite exciting since, 
for example, it now opens up a possibility for solving large-$N$ limit 
of ${\cal N}=4$ super-Yang Mills theory exactly.

As is now well-known, Type IIB string theory possesses 
$SL(2,{\bf Z})$-duality symmetry~\cite{schwarz10}, 
which is believed to hold not only for flat spacetime but also
for any classical background. Hence, it would be of some 
interest to check 
whether the D-brane actions on $AdS^5\times S^5$ has indeed the right 
transformation property under the $SL(2,{\bf Z})$-duality. 
In particular, on
the $AdS^5 \times S^5$ background, the D3-brane action 
should be self-dual 
under $SL(2,{\bf Z})$, and the actions for D1- and D5-branes 
should form a
multiplet of $SL(2,{\bf Z})$. 

In this paper, we prove explicitly the 
expected $SL(2,{\bf Z})$ transformation properties of the D1- and 
D3-brane
actions on $AdS_5 \times S^5$, at least in the semi-classical limit. 
The formal expression of 
$AdS_5 \times S^5$ D3-brane action obtained in~\cite{MeTs} is quite 
similar 
to the action in flat spacetime~\cite{Ag2}. This then suggests that 
the proof 
of the $SL(2,  {\bf Z})$ duality symmetry ought to be similar to that 
for the 
flat spacetime case~\cite{Ag}. An important ingredient 
in the proof of the
$SL(2,{\bf Z})$ 
symmetry in the flat spacetime has been that the $S$-generator of 
the $SL(2, {\bf Z})$ corresponding to 
$\tau \rightarrow -\frac{1}{\tau}$ acts as a 
finite $SO(2)$ rotation in the space of the two fermionic 
coordinates~\cite{
Ag, igarashi}. 
We show that the same continues to be true on $AdS_5 \times S^5$ 
background. 
In fact, the ${\cal N}=2$ Type IIB supersymmetry algebra from which the 
Mauer-Cartan equations are derived is covariant 
under the $SO(2)$ rotation. 
Stated differently, if a set of the two supercharges, $Q_{1}$ and 
$Q_{2}$, 
is related to an another set, $Q_1'$ and $Q_2'$, by a $SO(2)$ rotation, 
then 
the ${\cal N}=2$ supersymmetry algebra turns out to retain the same form 
when expressed in either set of the supercharges.

In section 2, we recapitulate relevant part of 
the construction~\cite{MeTs} of D3-brane action on $AdS_5 \times S^5$. 
We then prove 
the self-duality of the action 
under $SL(2, {\bf Z})$ by performing the duality transformation 
on the world-volume
fields as well as the background fields. The proof is in fact almost 
the same 
as that for the flat space case~\cite{Ag}. In section 3, 
we proceed to the 
D1-brane and construct the action on $AdS_5 \times S^5$. 
We then show that
the action is, as expected, $SL(2,{\bf Z})$ 
covariant~\footnote{The behavior under 
the $SL(2,{\bf Z})$ transformation in the general Type II background 
has been also discussed in~\cite{Ce2, CeTo}.}.

As our work is being finished, we have received related works~\cite{oda}.
In~\cite{oda}, gauge fixing of the $\kappa$-symmetry 
was indispensible for showing the $SL(2,{\bf Z})$ symmetry.
Our proof is based only on the covariance of the $SU(2,2 \vert 4)$ 
superalgebra under the
$SO(2)$-rotation and hence proves the $SL(2,{\bf Z})$ duality 
symmetry without
any gauge fixing. 

\medskip

\section{$SL(2,{\bf Z})$ Invariance of The D3-brane on $AdS_5\times S^5$}
  
The superstring or the D3-brane action can be constructed using the 
$SU(2,2|4)$ superalgebra. If we consider the coset superspace 
$\frac{SU(2,2|4)}{SO(1,4)\times SO(5)}$, the even part is 
$\frac{SO(2,4)}{SO(1,4)}\otimes \frac{SO(6)}{SO(5)}\simeq AdS^5\otimes 
S^5$. The corresponding algebra is described in $SO(1,4)\times SO(5)$ 
 basis, as shown in \cite{MeTs}. 
The even generators are two pairs of translation 
and rotation, $(P_a, J_{ab})$ for $AdS^5$ and $(P_a', J_{ab}')$ for 
$S^5$ and the odd generator are the two 10-d Majorana-Weyl 
spinors $Q^{\alpha\alpha'}_I$. 
Let $\gamma_a$ and $\gamma_a'$ be the $4\times 4$ matrices  
generating  the $SO(1,4)$ and $SO(5)$ Clifford algebra with 
the signature $\gamma^{(a}\gamma^{b)}=(-++++)$ and $\gamma^{(a'}
\gamma^{b')}=(+++++)$. The 10-dimensional $32\times 32$ Dirac 
matrices $\Gamma^{\hat{a}}$ of $SO(1,9)$ are represented as  
\begin{equation}
\Gamma^a=\gamma^a \otimes I\otimes \tau_1, \,\,\,\, 
\Gamma^{a'}=I\otimes \gamma^{a'}\otimes \tau_2
\end{equation}
where $I$ is the $4\times 4$ unit matrix and $\tau_i$ are 
the Pauli matrices. 
In \cite{MeTs}, the diagonal Majorana 
condition is chosen
\begin{equation}
\bar{Q}_{\alpha\alpha'I}\equiv 
(Q_I^{\beta\beta'})^{\dagger}(\gamma^0)_{\alpha}^{\beta}
\delta_{\alpha'}^{\beta'}=-Q_I^{\beta\beta'}C_{\beta\alpha}
C_{\beta'\alpha'}
\end{equation} 
where $C=(C_{\alpha\beta})$ and $C^{\prime}=(C_{\alpha'\beta'})$ 
are the charge conjugation matrix of the $SO(4,1)$ and $SO(5)$ 
Clifford algebra. 
The part of the supersymmetry algebra containing the spinors is 
\begin{eqnarray}
\left[ Q_I, P_a\right] &=&-\frac{i}{2}\varepsilon_{IJ}Q_J \gamma_a 
\nonumber \\
\left[ Q_I, P_a'\right] &=& \frac{1}{2}\varepsilon_{IJ}Q_J\gamma_a' 
\nonumber \\
\left[ Q_I, J_{ab} \right] &=& -\frac{1}{2}Q_I\gamma_{ab} \nonumber \\
\left[ Q_I, J_{a'b'}\right] &=& -\frac{1}{2}Q_I\gamma_{a'b'} \nonumber 
\end{eqnarray}
\begin{eqnarray}
\{ Q_{\alpha\alpha'I}, Q_{\beta\beta'J} \}  
 &=&\delta_{IJ}(-2iC_{\alpha'\beta'}(C\gamma^a)_{\alpha\beta}P_a
+2C_{\alpha\beta}(C'\gamma^{a'})_{\alpha'\beta'}P_{a'})  \nonumber \\ 
 & & +\varepsilon_{IJ}(C_{\alpha'\beta'}(C\gamma^{ab})_{\alpha\beta}J_{ab}
-C_{\alpha\beta}(C'\gamma^{a'b'})_{\alpha'\beta'}J_{a'b'}). 
\end{eqnarray}
Note that the diagonal Majorana condition is preserved under 
the $SO(2)$ rotation between $Q_1$ and $Q_2$. And one can easily 
check that the $SU(2,2|4)$ superalgebra is covariant under the $SO(2)$ 
rotation, as $\delta_{IJ}$ and $\varepsilon_{IJ}$ are invariant 
under $SO(2)$. 

The left invariant Cartan 1-forms $L^A=dX^ML^{A}_{M}, \, X^M=(x,\theta)$
are given by 
\begin{equation}
G^{-1}dG=L^AT_A=L_aP^a+L_{a'}P^{a'}
+\frac{1}{2}L^{ab}J_{ab}+\frac{1}{2}L^{a'b'}J_{a'b'}
+L^{\alpha\alpha'I}Q_{\alpha\alpha'I}
\end{equation}
where $G=G(x,\theta)$ is a coset representative in $SU(2,2|4)$, 
$L^a, L^{a'}$ are the 5-beins,  $L^{\alpha\alpha'I}$ are two 
spinor 16-beins and $L^{ab},L^{a'b'}$ are the Cartan connection.

The D3-brane action 
presented in \cite{MeTs} is\footnote{Our convention is that whenever 
an integral without $d^n\sigma$, 
it is an integral of a differential form.}
\begin{equation}
S = - \int d^4 \sigma  \sqrt{-{\rm det}\, (G_{ij}+
e^{-\frac{\phi}{2}} F_{ij}-b_{ij})}
+ \int \left(C_4+C_2\wedge (e^{-\frac{\phi}{2}} F-b_2) 
+\frac{1}{2} C_0 F\wedge F \right) \, . \label{eq:d3}  
\end{equation}
Here we have $G_{ij}=L_i^{\hat{a}}L_j^{\hat{a}}=\partial_iX^M
L_{M}^{\hat{a}}\partial_j X^N L_N^{\hat{a}}$ where $\hat{a}=(a,a')$ 
run through the $SO(1,4)$ and $SO(5)$ indices and $L^{\hat{a}}(X(\sigma))
=d\sigma^iL_i^{\hat{a}}$. Also $b_2\equiv 
\frac{1}{2}b_{ij} d\sigma^i d\sigma^j $ 
and $db_2=i\bar{L}\tau_3 \hat{L}L$ 
with $\hat{L}_i=L_i^{\hat{a}}\Gamma^{\hat{a}}$. 
Finally $C_4$ and $C_2$
 are determined by the condition 
\begin{eqnarray}
H\equiv d(C_4+C_2\wedge (e^{-\frac{\phi}{2}} F-b_2)
&=&\frac{i}{6}\bar{L}\hat{L}^3\tau_3\tau_1L
+i\bar{L}\tau_1(e^{-\frac{\phi}{2}} F-b_2)\hat{L} L \nonumber \\
& &+\frac{1}{30}(\varepsilon^{a_1...a_5}L^{a_1}\wedge \cdots \wedge 
L^{a_5}+\varepsilon^{a'_1 ...a'_5}L^{a'_1}\wedge \cdots \wedge 
L^{a'_5}).  \label{eq:c4}
\end{eqnarray} 
From this condition one obtains the following useful identity:
\begin{equation}\label{eq:id3}
dC_4-C_2\wedge db_2=
\frac{i}{6}\bar{L} \tau_3\tau_1 \hat{L}^3 L
 +\frac{1}{30}(\varepsilon^{a_1...a_5}L^{a_1}\wedge \cdots \wedge 
L^{a_5}+\varepsilon^{a'_1 ...a'_5}L^{a'_1}\wedge \cdots \wedge 
L^{a'_5}).  
\end{equation}
In Eq.~(\ref{eq:d3}), for later convenience, we have written down 
the dilaton 
dependence in the Einstein frame of the $AdS_5 \times S^5$ background. 
In fact, the string and Einstein frames are equivalent for D3-brane, 
since 
the dilaton takes a constant value on $AdS_5 \times S^5$. 
The $C_0$ term in Eq.~(\ref{eq:d3})
is absent in \cite{MeTs}, but it is a total 
derivative term (or boundary term)
that can be added to the action without changing the 
classical equations of motion. The inclusion of the constant $C_0$ 
is consistent since the dilaton remains constant in the 
$AdS^5\times S^5$ background. In fact, a constant shift of $C_0$ is 
a trivial classical symmetry of the action. 
Inclusion of the $C_0$ term, however, is quite crucial for the proof
since the $SO(2)$ duality symmetry is promoted 
to the full $SL(2,{\bf R})$  
symmetry in the presence of both dilaton and $C_0$ field background.

Adding a Lagrange multiplier term $\int d^4\sigma \frac{1}{2}
\tilde{H}^{ij}\, [F_{ij}-(\partial_iA_j - \partial_j A_i) ]$ 
to the action Eq.~(\ref{eq:d3}), the $A_i$ equation of motion 
can be solved by 
$\tilde{H}^{ij} =\varepsilon^{ijkl}\partial_{k}B_l$. 
The duality transformation is essentially the same as 
in the flat case~\cite{Ag} and the resulting action is 

\begin{equation} 
S_D=  - \int d^4 \sigma  \sqrt{-{\rm det}\, \left(G_{ij}
+\frac{1}{\sqrt{1+e^{2\phi}C_0^{ 2}}} (e^{\frac{\phi}{2}}
\tilde{F}_{ij}
+C_{ij}+e^{\phi}C_0 b_{ij} \right)}
+\int \Omega_D,  
\end{equation}
where
\begin{eqnarray}
 \Omega_D&=& C_4-b_2\wedge C_2-\frac{1}{2}e^{\phi}C_0  
b_2\wedge b_2
+b_2\wedge \left(e^{\frac{\phi}{2}}\tilde{F} +C_2+e^{\phi}C_0  b_2 \right)
  \nonumber  \label{eq:lw2} \\ 
 & &-\frac{e^{\phi}C_0 }{2(1+e^{2\phi}C_0^{ 2})}
\left(e^{\frac{\phi}{2}}\tilde{F} +C_2+e^{\phi}C_0  b_2 \right)
\wedge \left(e^{\frac{\phi}{2}}\tilde{F} +C_2+e^{\phi}C_0  b_2 \right)
\end{eqnarray}
where $\tilde{F}=dB$.
If we define rotated $SO(2)$ Pauli matrices 
\begin{eqnarray}
\tau_1^{\prime} &\equiv& - \left({\tau_3 +e^{\phi}C_0\tau_1} \right)
/{\sqrt{1+e^{2\phi}C_0^2}} 
\nonumber \\
\tau_3^{\prime} &\equiv& + \left({\tau_1-e^{\phi}C_0\tau_3} \right)
/{\sqrt{1+e^{2\phi}C_0^2}} \label{eq:rot}
\end{eqnarray}
and using Eq.~(\ref{eq:id3}) and the identity $\tau_3\tau_1=\tau_3'\tau_1'$ , 
we were able to show that 
\begin{equation}
d\Omega_D=\frac{i}{6}\bar{L}\hat{L}^3\tau_3'\tau_1'L
+i\bar{L}\tau_1'(\frac{e^{\frac{\phi}{2}}}{\sqrt{1+e^{2\phi}C_0^2}}
\tilde{F}-b_2')\hat{L}L
+\frac{1}{30}(\varepsilon^{a_1...a_5}L^{a_1}\wedge \cdots \wedge 
L^{a_5}+\varepsilon^{a'_1 ...a'_5}L^{a'_1}\wedge \cdots \wedge 
L^{a'_5}) \, .  
\end{equation}

Thus the dual action can be written as 
\begin{eqnarray}
S&=& -\int d^4 \sigma \sqrt{-{\rm det}\, \left(G_{ij}
+\frac{e^{\frac{\phi}{2}}}{\sqrt{1+e^{2\phi}C_0^2}} \tilde{F}_{ij}
-b'_{ij} \right)} \nonumber \\
& &+\int \left\{ C_4' + C_2' \wedge \left(\frac{e^{\frac{\phi}{2}}}
{\sqrt{1+e^{2\phi}C_0^2}}\tilde{F} - b'_2 \right)
-\frac{e^{2\phi}C_0}{2(1+e^{2\phi}C_0^2)}\tilde{F}\wedge \tilde{F} 
\right\} \, .
\label{eq:dd3}
\end{eqnarray}
where the tensor fields $C_4',C_2'$ and $b_2'$ are the same as 
$C_4,C_2$ and $b_2$ but with $SO(2)$ rotated Pauli matrices in the 
corresponding expression.
The resulting action still has the $\kappa$-symmetry since the Mauer-Cartan 
equation and Fierz identity used in the derivation of the $\kappa$-symmetry 
are covariant under the $SO(2)$ rotation. 

Also, we can check the transformation of the dilaton and 
the axion under the duality transformation. 
{}From the coefficient of $\tilde{F}$ in the Born--Infeld part of 
Eq.~(\ref{eq:dd3}), we obtain the transformation
\begin{equation}
e^{-\phi} \qquad \rightarrow \qquad + \frac{e^{\phi}}{1+e^{2\phi}C_0^{2}}
=\frac{1}{e^{\phi}+e^{-\phi}C_0^{2}} \, . 
\end{equation}
From the coefficient of $\tilde{F} \wedge \tilde{F}$  
, we also have 
\begin{equation}
\,\,\, C_0 \qquad \rightarrow \qquad -\frac{e^{2\phi}C_0}
{1+e^{2\phi}C_0^{2}}=-\frac{e^{\phi}C_0 }
{e^{-\phi}+e^{\phi}C_0^{2}} \, .
\end{equation}
Combining this transformation with the symmetry under a 
constant shift of $C_0$, at the classical level, one deduces that the D3-brane 
action has $SL(2,{\bf R})$ symmetry. Once semi-classical quantum effects on the
D3-brane world-volume are taken into account, one finds that magnetic monopoles
should be introduced as ends of the D-strings. In this case, the constant
shift of $C_0$ is replaced by a quantized shift, much as in QCD. 
Thus, the $U(1)$ gauge group is compact, and the surviving symmetry group of 
quantum Type IIB string theory is $SL(2, {\bf Z})$.

\medskip

\section{$SL(2,{\bf Z})$ Covariance of The D1-brane on $AdS_5 \times S^5$}

We now turn to the D1-brane on $AdS_5 \times S^5$.
The action of the D1-brane with $\kappa$-symmetry is given by
\begin{equation}
S = - \int d^2 \sigma e^{-\phi} \sqrt{-{\rm det}\, (G_{ij}
+ F_{ij}-b_{ij})} + \int e^{-\phi} C_2 \, .
\label{eq:sd1}
\end{equation}
where $C_2$ is determined by $dC_2=i\bar{L}\tau_1\hat{L}L$. 
We were able to check explicitly, using Mauer-Cartan equation and 
Fierz identity, that $dC_2$ is closed. Alternatively, we can make a $SO(2)$ 
rotation mapping $\tau_1$ to $\tau_3$. The resulting 3-form 
is the standard closed 3-form appearing in the superstring action 
\cite{MeTs2}. In Eq.~(\ref{eq:sd1}), we have expressed the dilaton dependence 
in the string frame. The $\kappa$-symmetry transformation is given by 
\begin{equation}
\delta_{\kappa}x^{\hat{a}}=0, \,\, \delta_{\kappa}\theta=\kappa
\, , 
\label{eq:k1}
\end{equation}
where the $\kappa$-parameter satisfies
\begin{equation}
\Gamma \kappa=\kappa, \, \,\, \Gamma^2=1  \label{eq:k2}
\, 
\end{equation}
and the matrix $\Gamma$ is given by 
\begin{equation}
\Gamma=\frac{\varepsilon^{i_1i_2}}{\sqrt{-det(G_{ij}+F_{ij}-b_{2ij})}}
\left(\frac{1}{2}\tau_1\hat{L}_{i_1}\hat{L}_{i_2}
+\frac{i\tau_2}{2}(F_{i_1i_2}-b_{2i_1i_2}) \right) \, .
\end{equation} 
The corresponding variation of $G_{ij}$ and $F_{ij}-b_{ij}$ is 
given by 
\begin{eqnarray}
\delta_{\kappa}G_{ij}&=&-2i\delta_{\kappa}\bar{\theta}
(\hat{L}_i L_j +\hat{L}_j L_i) \label{eq:k3}\\
\delta_{\kappa}(F-b_2)&=&+ 2i\delta_{\kappa}\bar{\theta}\hat{L}\tau_3 L \, .
\end{eqnarray}
One can also add a total derivative term 
\begin{equation}
e^{-\phi} C_2 \qquad \rightarrow \qquad e^{-\phi} C_2 -C_0 F,
\end{equation}
where $C_0$ is a constant ``axion'' background field.
Since $C_0 F$ is a total derivative, 
it does not affect the classical equations
of motion. As in the previous section, a constant shift of $C_0$ is a trivial 
classical symmetry of the action Eq.~(\ref{eq:sd1}). At quantum level, the 
symmetry is reduced to a discrete shift symmetry, $T$-symmetry of 
$SL(2, {\bf Z})$.

Now we perform the duality transformation. 
We first introduce a Lagrange multiplier field
$\tilde{H}^{\mu\nu}=-\tilde{H}^{\nu\mu}$ as follows:
\begin{equation}
S^{\prime}= \int d^2 \sigma \left(- e^{-\phi} \sqrt{-{\rm det}\, 
(G_{ij}+
F_{ij}-b_{ij})}
+\frac{1}{2} \tilde{H}^{ij} (F_{ij} - 2 \partial_{i} A_{j} )
+  \frac{1}{2} e^{-\phi}\epsilon^{ij}C_{ij}
-\frac{1}{2}C_0 \epsilon^{ij} F_{ij} \right)
 \label{eq:sd2}
\end{equation}
and regard the field-strength $F_{ij}$ to be an independent
field.
Varying the action with respect to $A_{j}$ yields 
$\partial_{i}\tilde{H}^{ij}=0$,
which implies that $\tilde{ H}^{ij}=\epsilon^{ij}
\Lambda$ with $\Lambda$ constant.
If we use the equation of motion for $F$ to rewrite
the action in terms of $\Lambda$ instead of $F$,
 the total action can be written as
\begin{equation}
S_D= \sqrt{e^{-2\phi}+(\Lambda-C_0)^2}
\int d^2 \sigma \left(-\sqrt{-{\rm det}\, G_{ij}}
-\frac{1}{2}\epsilon^{ij}b_{ij}^{\prime} \right)  \label{eq:S1d}
 \end{equation}
where $db_2^{\prime}=i\bar{L}\tau_3^{\prime}\hat{L}L$
and $\tau_3'$ is defined to be 
\begin{equation}
e^{-\phi} \, \tau_1-(\Lambda-C_0) \, \tau_3\equiv
\sqrt{(e^{-2\phi}+(\Lambda-C_0)^2)} \,  \tau_3' \, .
\end{equation} 
The action Eq.~(\ref{eq:S1d}) is  the $\kappa$-symmetric
superstring action in the Nambu-Goto form with the modified tension
\begin{equation}
T^{\prime}=\sqrt{e^{-2\phi}+(\Lambda-C_0)^2} \, . \label{eq:tension}
\end{equation}
The $\kappa$-transformation is given by Eqs.~(\ref{eq:k1}), (\ref{eq:k2}),
(\ref{eq:k3}), but now $\Gamma$ is given by
\begin{equation}
\Gamma=\frac{\varepsilon^{i_1i_2}}{\sqrt{-det(G_{ij})}}
(\frac{1}{2}\tau_3\hat{L}_{i_1}\hat{L}_{i_2}) \, .
\end{equation}

This result admits the same interpretation as in the flat space. 
The expression Eq.~(\ref{eq:tension}) can be interpreted as the tension 
for the $SL(2, {\bf Z})$ covarinat spectrum of
strings provided that one identifies the integer value $\Lambda = m$
as corresponding to the $SL(2,{\bf Z})$ dyonic string of charge $(m, 1)$ 
in the $AdS_5 \times S^5$ background 
with constant dilaton $\phi$ and axion $C_0$. An
equivalent interpretation is that Eq.~(\ref{eq:S1d}) describes 
the fundamental $
(1,0)$
string with an $SL(2, {\bf Z})$ transformed metric, dilaton and axion.
The relevant $SL(2, {\bf Z})$ transformations maps $C_0+ie^{-\phi}$ to
$-(C_0-\Lambda+ie^{-\phi})^{-1}$. Thus the coupling constant of the
fundamental string after the duality transformation is given by
$e^{\tilde{\phi}}=e^{-\phi}+e^{\phi}(\Lambda-C_0)^2$.

\medskip

\section{Discussion}

In this paper, using coset superspace approach, we have investigated the 
$SL(2, {\bf Z})$ duality property of the D1-and D3-brane actions 
on $AdS^5\times S^5$ background. We have proven that the D1-and D3-branes 
behave as in the expected way under the $SL(2, {\bf Z})$ transformation. 
Moreover, we have shown that the actual dual transformation is 
formally the 
same as in the flat space. 
In both cases, the $SL(2, {\bf Z})$ acts on the fermionic 
coordinates in the 
same way, viz, as the $SO(2)$ rotation. 
In particular, with the dilaton and axion field background turned on, 
we have shown that the full $SL(2, {\bf Z})$ duality symmetry 
follows from 
the $SO(2)$ duality rotation. 
Throughout this paper, our discussion has been mainly semi-classical. 
It would be interesting 
to extend our proof to the full-fledged quantum level. 
\vskip0.5cm
S.J.R. thanks Institute for Advanced Study and Aspen Center for
Physics for hospitality, where part of this work was carried out.

\end{document}